\documentclass{article}
\usepackage[preprint]{spconfa4}
\usepackage{amsmath,graphicx,url,times,amsmath,amssymb,acronym,balance}
\usepackage{color,booktabs}
\usepackage[table]{xcolor}
\usepackage{cite}
\usepackage{lipsum}

\copyrightnotice{978-1-6654-6867-1/22/\$31.00~\copyright2022 IEEE}
\toappear{To appear in {\it Proc.\ IWAENC2022, Sept. 05-08, 2022, Bamberg, Germany}}

\title{Task splitting for DNN-based acoustic echo and noise removal}
\name{Sebastian Braun, Maria Luis Valero} 
\address{
	Microsoft Corporation, USA \\
	\{sebastian.braun, maria.luis\}@microsoft.com}
\acrodef{STFT}{short-time Fourier transform}
\acrodef{MSE}{mean-squared error}
\acrodef{MAE}{mean absolute error}
\acrodef{PSD}{power spectral density}
\acrodef{RTF}{relative transfer function}
\acrodef{SNR}{signal-to-noise ratio}
\acrodef{segSNR}{segmental signal-to-noise ratio}
\acrodef{SRR}{signal-to-reverberation ratio}
\acrodef{PDF}{probability density function}
\acrodef{DOA}{direction-of-arrival}
\acrodef{VAD}{voice activity detector}
\acrodef{MVDR}{minimum variance distortionless response}
\acrodef{AIR}{acoustic impulse response}
\acrodef{PESQ}{perceptual evaluation of speech quality}
\acrodef{STOI}{short-time objective intelligibility}
\acrodef{LSD}{log spectral distance}
\acrodef{CD}{cepstral distance}
\acrodef{WER}{word error rate}
\acrodef{SPP}{speech presence probability}
\acrodef{DNN}{deep neural network}
\acrodef{RNN}{recurrent neural network}
\acrodef{CNN}{convolutional neural network}
\acrodef{FC}{fully connected}
\acrodef{CRN}{convolutional recurrent network}
\acrodef{LSTM}{long-term short-term}
\acrodef{GRU}{gated recurrent unit}
\acrodef{FF}{feed forward}
\acrodef{ReLU}{rectified linear unit}
\acrodef{GCC}{generalized cross-correlation}
\acrodef{RMSE}{root-mean-square error}
\acrodef{CPSD}{cross-power spectral density}
\acrodef{siSDR}{scale-invariant signal-to-distortion ratio}
\acrodef{SDR}{signal-to-distortion ratio}
\acrodef{MagMSE}{Magnitude MSE}
\acrodef{LPS}{logarithmic power spectrum}
\acrodef{CSE}{complex spectrum error}
\acrodef{MagSE}{magnitude spectrum error}
\acrodef{LogMagSE}{logarithmic magnitude spectrum error}
\acrodef{DL}{deep learning}
\acrodef{PLSD}{phase-aware logarithmic spectrum distance }
\acrodef{SDW}{speech distortion-weighted}
\acrodef{MOS}{mean opinion score}
\acrodef{RIR}{room impulse response}
\acrodef{MAC}{multiply-accumulate}
\acrodef{CRUSE}{Convolutional Recurrent U-net for Speech Enhancement}
\acrodef{FD}{frequency-domain}
\acrodef{DNS}{deep noise suppression}
\acrodef{SED}{sound event detection}
\acrodef{ASR}{automatic speech recognition}
\acrodef{ERB}{equivalent rectangular bandwidth}
\acrodef{RSNR}{reverberant speech-to-noise ratio}
\acrodef{AEC}{acoustic echo cancellation}
\acrodef{EIR}{echo impulse response}
\acrodef{CCMSE}{complex compressed mean-squared error}
\acrodef{SER}{signal-to-echo ratio}
\acrodef{NRES}{noise and residual echo suppression}
\acrodef{NS}{noise suppression}
\acrodef{MSC}{magnitude-squared coherence}
\acrodef{DAEC}{deep AEC}

\definecolor{matlab1}{rgb}{0, 0.4470, 0.7410}
\definecolor{matlab2}{rgb}{0.8500, 0.3250, 0.0980} 
\definecolor{matlab3}{rgb}{0.9290, 0.6940, 0.1250} 
\definecolor{matlab4}{rgb}{0.4940, 0.1840, 0.5560} 
\definecolor{matlab5}{rgb}{0.4660, 0.6740, 0.1880}  

\begin{document}
	\ninept
	\maketitle
	\begin{abstract}
		Neural networks have led to tremendous performance gains for single-task speech enhancement, such as noise suppression and acoustic echo cancellation (AEC). In this work, we evaluate whether it is more useful to use a single joint or separate modules to tackle these problems. We describe different possible implementations and give insights into their performance and efficiency.
		We show that using a separate echo cancellation module and a module for noise and residual echo removal results in less near-end speech distortion and better performance during double-talk at same complexity.
	\end{abstract}
	\section{Introduction}
	\label{sec:intro}
	The main acoustic problems tackled by typical speech communication pipelines are removal of echo, noise, and reverberation, which in many cases occur simultaneously \cite{Hansler2004}. The adoption of \ac{DL} for real-time speech enhancement 
	has made fast progress over the last few years: neural networks have been developed at small enough size for practical applications \cite{Tan2020,Fedorov2020,Braun2021a} far outperforming traditional \ac{NS} techniques. A more recent research spike on \ac{AEC}, in part fueled by the Microsoft AEC challenge series \cite{Cutler2021}, shows strong trends and promising results adopting neural networks for \ac{AEC}. However, while joining rather separate tasks, such as \ac{NS} and \ac{AEC}, into a single model became very straightforward using \ac{DL}, it is not yet understood at what cost this comes: How much larger does a joint \ac{NS}+\ac{AEC} model have to be to achieve on par \ac{NS} performance to an \ac{NS}-only model? Is it more efficient to break the tasks into separate stages, or is \ac{DL} so powerful that simply training an unconstrained single-stage black-box model will yield better performance or higher efficiency?

	In \cite{Carbajal2020}, a system for joint reverberation, noise and echo reduction using a {DNN} supported EM algorithm is proposed. In \cite{Halimeh2019,Ivry2021} it was proposed to use a linear \ac{AEC} system extended with a \ac{DNN} to cope with non-linear components.
	Most straightforward full \ac{DNN} based systems use spectral mapping \cite{Zhang2019,Peng2021}, or predict spectral enhancement filters or masks \cite{Westhausen2021,Valin2021}. Many participants in the AEC challenge \cite{Cutler2021} used a hybrid system with a linear \ac{AEC} and a \ac{DNN} module.
	In \cite{Franzen2021,Seidel2021} a two-stage approach was proposed that uses a dedicated AEC module and a second \ac{NRES} module.
	In \cite{Li2021}, a multi-stage enhancement system with separate dereverberation and denoising modules was proposed. While the authors showed the subsequent performance gained by each stage, including better performance than other baselines, it was not proven if the target decoupling has an actual benefit over an unconstrained optimization of a similar network. Often, such multi-stage processing approaches are also trained sequentially, requiring a training process for each stage.
	\cite{Ivry2021a} proposes a tunable loss function to increase echo suppression at the cost of speech distortion.
	
	In this paper, we design a two-stage \ac{DNN} based system consisting of \ac{DAEC} and \ac{NRES} modules. We propose an adaptive loss that avoids burdensome multi-stage training.
	The two-stage system is compared to fair single-stage \acp{DNN} trained on the echo and noise suppression task. We show that this way, the \ac{AEC} module is removing only echo, which creates no significant signal distortion in contrast to echo and noise suppressors. The \ac{NRES} module is cleaning the signal up, removing eventual residual echoes and noise, while introducing only moderate amounts of signal distortion. Overall, we show that the proposed two-stage system outperforms the single-stage baseline in terms of signal distortion and double-talk.

	\section{Problem formulation}
	\label{sec:problem}
	Given a typical full-duplex communication system comprising a loudspeaker and microphone in the same room, we assume the following signal arriving at the microphone
	\begin{equation}
		\label{eq:sigmodel}
		y(t) = s(t) + r(t) + n(t) + \underbrace{h(t) \star \mathcal{Q}\{u(t)\}}_{d(t)},
	\end{equation}
	where $s(t)$ is the desired speech signal, which may also include early reflections, $r(t)$ is the (late) reverberation, $n(t)$ is additive noise, $u(t)$ is the far-end signal, $h(t)$ is the \ac{EIR} that describes the propagation from the loudspeaker to the microphone, $\mathcal{Q}$ denotes a nonlinear function modeling e.g.\ loudspeaker and possible processing distortions, and $t$ is the time index.
	
	We use capital letters to denote \ac{STFT} representations of the time-domain signals, e.\,g.\ $Y(k,n)$ is the \ac{STFT} of $y(t)$ with frequency and time indices $k,n$. Typical speech enhancement systems estimate a time-frequency filter 
	to map the input signal spectrum $Y(k,n)$ to an estimate of the desired speech signal $\widehat{S}(k,n)$.
	In this work, we use a generalized convolutive cross-band filter $G_{k,n}(\kappa,\ell)$ taking also neighboring frames and frequencies for the mapping per time-frequency point into account. In \cite{Mack2019} this was called \emph{deep filtering} and is described by
	\begin{equation}
		\label{eq:mic-enhancement}
		\widehat{S}(k,n) = \sum_{\kappa=-K}^K \sum_{\ell=0}^L G_{k,n}(\kappa,\ell) \, Y(k-\kappa, n-\ell),
	\end{equation}
	where $K$ are the number of neighboring frequencies and $L$ the number of past frames used. The filter is therefore causal.
	The time-domain counterpart $\widehat{s}(t)$ is obtained by inverse \ac{STFT}.

	\section{Model architecture}
	We base our models on the \ac{CRUSE} architecture \cite{Braun2022} with the only modification of using the deep filter \eqref{eq:mic-enhancement} instead of single time-frequency mapping with $K\!=\!L\!=\!0$. Our CRUSE configuration uses 4 convolutional encoder layers, mirrored transposed-convolutional decoder layers with $[32,64,64,64]$ channels, causal kernels of (2,3) in dimensions (time,frequency), and downsampling along the frequency axis using strides (1,2). Between encoder and decoder is a grouped \ac{GRU} of 4 groups to reduce complexity \cite{Tan2020,Braun2021a}. From each encoder, a skip connection with a 1x1 convolution is added to the input of each corresponding decoder layer.

	\subsection{Single-stage model}
	As baseline and single-stage model, we use the CRUSE-NS model as described in \cite{Braun2022}, with the only modification of the deep filtering \eqref{eq:mic-enhancement}. To deal with the task of \ac{AEC}, the only modification to obtain the baseline single-stage CRUSE-AEC model is doubling the input channels of the convolutional encoder from 2 to 4 to process real and imaginary parts of mic and far-end signals. As input features, the complex spectra are compressed as in \cite{Li2021a,Braun2022}. The single-stage adapted model is shown in Fig.~\ref{fig:twostage-diag}a).
	
	The far-end signal is delay-aligned to the mic signal in the \ac{STFT} domain on frame basis using a simple \ac{MSC} based delay estimation. The delay is found as the frame with the highest smoothed \ac{MSC} between mic and far-end signal on a 1~s window without using look-ahead in online processing fashion. We found \ac{MSC} based alignment to be more robust and less complex than time-domain based correlation methods.
	
	\subsection{Task splitted two-stage model}
	\begin{figure}[tb]
		\centering
		\includegraphics[width=0.8\columnwidth,clip,trim=230 260 190 180]{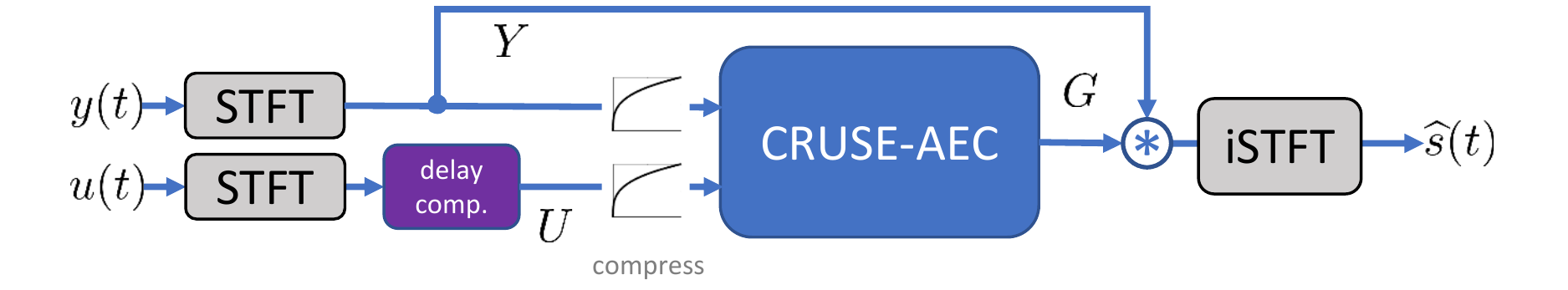}
		\newline a)
		\includegraphics[width=\columnwidth,clip,trim=110 160 185 150]{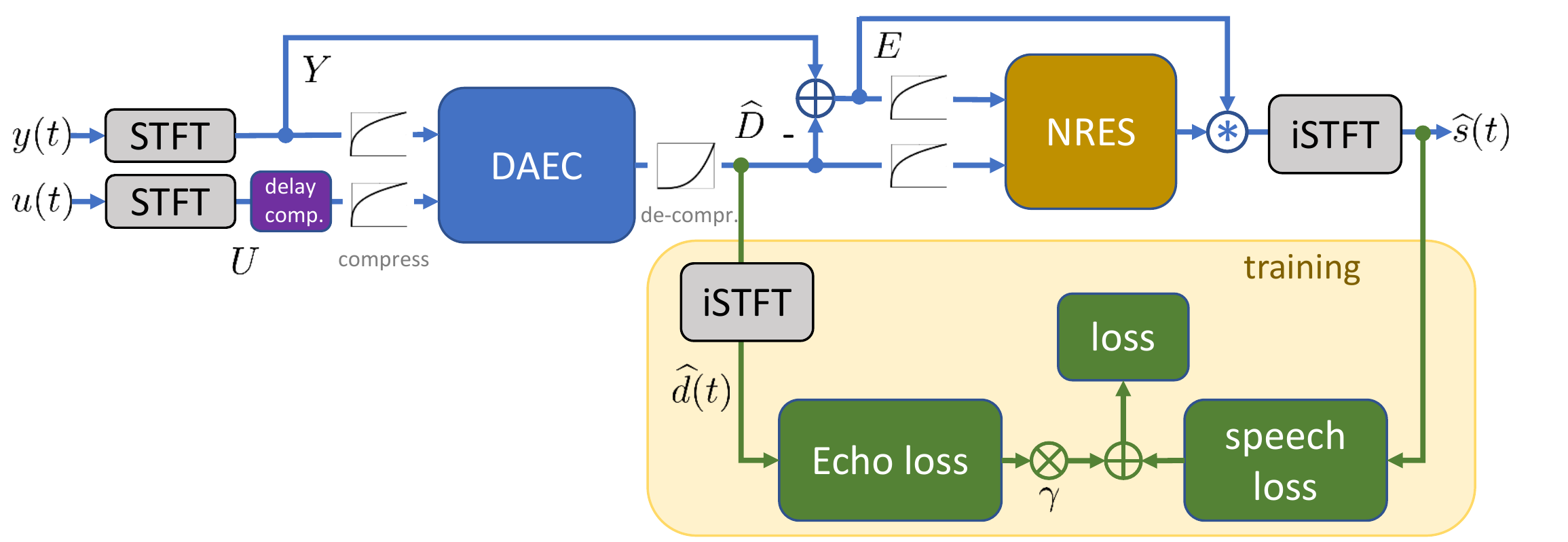}
		b)
		\caption{a) Single-stage CRUSE structure for AEC. b) Two-stage structure with deep AEC and NRES blocks and training strategy.}
		\label{fig:twostage-diag}
	\end{figure}
	The proposed two-stage AEC+NRES model structure is shown in Fig.~\ref{fig:twostage-diag}b). The AEC block estimates the compressed complex spectrum of the echo signal $d(t)$. We found the direct signal estimation to perform better than estimating a multi-frame filter for the far-end signal. 
	The echo signal is de-compressed and subtracted from the mic input by 
	\begin{equation}
		\label{eq:aec_out}
		E(k,n) = Y(k,n) - |\widehat{D}(k,n)|^{\frac{1}{c}} \, e^{j\varphi_{\widehat{D}}(k,n)},
	\end{equation}
	where $\varphi_{\widehat{D}}$ denotes the phase of $\widehat{D}$.
	The input to the NRES block, is the AEC output $E(k,n)$ and the echo estimate $\widehat{D}(k,n)$ as proposed in \cite{Seidel2021}, again with magnitude compression applied. Real and imaginary parts are fed as channels to the convolutional encoders, so both AEC and NRES modules have a 4-channel input.
	For training and analysis purposes only, also the echo estimate $\widehat{d}(t)$ is obtained via inverse \ac{STFT}.
	\begin{figure}[tb]
		\centering
		\includegraphics[width=0.8\columnwidth,clip,trim=240 235 350 190]{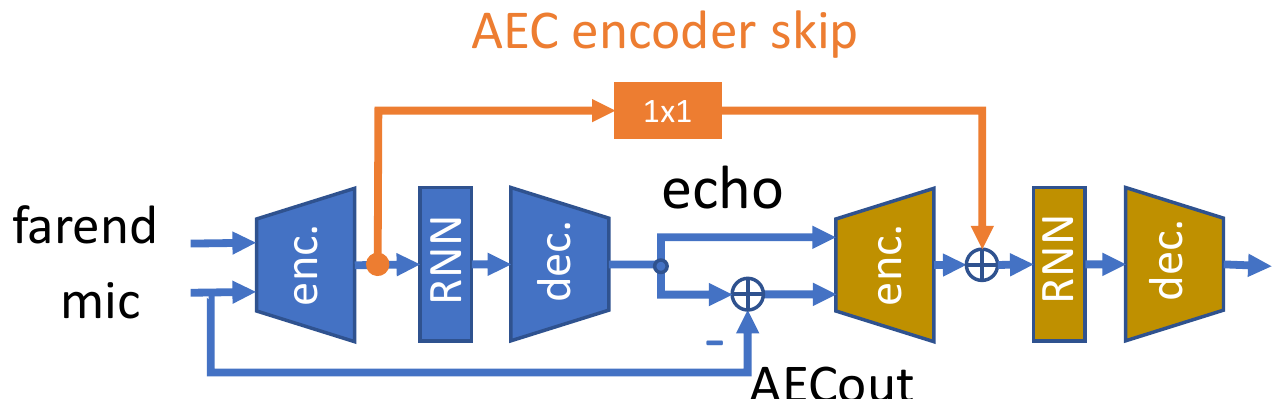}
		\vspace{-.2cm}
		\caption{Skip connection between DAEC and NRES module.}
		\label{fig:skip-arch}
	\end{figure}
	For better communication and re-use of the encoded far-end signal, we add a skip connection with $1\!\!\times\!\!1$ convolution from the AEC encoder to the end of the NRES encoder as shown in Fig.~\ref{fig:skip-arch}.


	\section{Loss function}
	As main loss component we aim to optimize the desired speech signal spectrum in an end-to-end fashion. We use the spectral \ac{CCMSE} loss, which outperformed other losses in a noise suppression task \cite{Braun2021b,Braun2022}. The \ac{CCMSE} loss is a linear combination of complex and magnitude components:
	\begin{equation}
		\label{eq:speechloss}
		\mathcal{L}(\widehat{s}, s) = \sum_{k,n} \alpha \left| |\widehat{S}|^c e^{j\varphi_{\widehat{S}}} - |S|^c e^{j\varphi_S} \right|^2 + (1\!-\!\alpha) \left||\widehat{S}|^c \!-\! |S|^c \right|^2
	\end{equation}
	
	Similarly as in \cite{Seidel2021}, we utilize a dedicated echo loss term to guide the AEC output to contain echo only. We found in preliminary experiments that using compressed spectral distances similar to \eqref{eq:speechloss} for the echo component results in significant under-estimation of the echo. Therefore, we propose to use the \ac{MAE}, which provides accurate echo estimates:
	\begin{equation}
		\label{eq:echoloss}
		\mathcal{L}(\widehat{d}, d) = \sum\nolimits_{k,n} \left|\widehat{D} - D \right|.
	\end{equation}
	Additionally, we use an asymmetric speech over-suppression penalty term \cite{Wang2020a}
	\begin{equation}
		\label{eq:asymloss}
		\mathcal{L}_\text{asym}(\widehat{s}, s) = \sum\nolimits_{k,n} \max \left\{ |S|^c - |\widehat{S}|^c ,\, 0 \right\}^2.
	\end{equation}
	The final loss is then given by the linear combination
	\begin{equation}
		\label{eq:overall_loss}
		\mathcal{L} = \mathcal{L}(\widehat{s}, s) + \beta \mathcal{L}_\text{asym}(\widehat{s}, s) + \gamma \mathcal{L}(\widehat{d}, d),
	\end{equation}
	where $\beta$ and $\gamma$ are positive scalars to balance the loss terms.
	The training loss strategy is outlined in Fig.~\ref{fig:twostage-diag} by the green blocks.

	\subsection{Proposed adaptive echo loss}
	To guide the training to first focus on providing a good echo estimate from the AEC module, we propose an adaptive weighting to combine end-to-end speech loss \eqref{eq:speechloss} and the echo term \eqref{eq:echoloss} with
	\begin{equation}
		\label{eq:res_echo_weight}
		\gamma = \max\left\{ \eta \, \frac{\sum_{k,n} \left|\widehat{D} - D \right|}{\sum_{k,n} |D|}, \quad\gamma_\text{min} \right\}
	\end{equation}
	where $\eta$ is a constant scalar and $\gamma_\text{min}\!>\!0$ prevents vanishing of the echo term.
	The adaptive weighting steers the training focus on the AEC module when the residual echo is large, and approaches zero when the AEC module cancels the echo sufficiently. Once this occurs, the training focuses on the \ac{NRES} module.

	\section{Training data and augmentation}
	We use supervised training by generating synthetic training signal mixtures, which provides access to individual signal components such as the desired speech training target and echo signals. We create training signals of 20~s length as given by the signal model in \eqref{eq:sigmodel}. The near-end speech signal is concatenated from speech recordings as described in \cite{Braun2022} using data from the \ac{DNS} challenge \cite{Reddy2021a} and AVspeech \cite{Ephrat2018}.
	The speech signal is convolved with a \ac{RIR} from the 115~k database provided in \cite{Reddy2021a}. The speech training target $s(t)$ is obtained by convolving the speech with a windowed part of the \ac{RIR} containing only early reflections up to 50~ms. The later part of the reverberation is considered as undesired signal component $r(t)$. 
	Noise signals are taken from the DNS challenge database (180~h). 80\% of the speech and noise signals are randomly modified in spectral shape, and 20\% in pitch. The noise is added to the speech with a \ac{SNR} with distribution $\mathcal{N}(5,10)$~dB. 
	
	The far-end signals are taken from three sources: noisy speech recordings (VoxCeleb2 \cite{Chung2018}, 4000~h), clean speech (650~h of VoxCeleb2 processed by noise suppression \cite{Braun2022}), and music from MUSAN \cite{Snyder2015} (42~h). Random signal portions from these databases are concatenated to the desired 20~s far-end signal length. The far-end is modified by inserting random silence periods of $[3,15]$~s, short audio drop-outs (10\%), and simulating clock drift by resampling with a standard deviation of 0.5 samples/sec (20\%). To model loudspeaker non-linearities, with 20\% chance sigmoid or rectifier clipping functions with random parameters are applied to the far-end signal.
	
	To generate the echo signal, one or multiple \acp{EIR} are chosen from the pool or \acp{RIR}. To simulate that all sources are in the same room, the \ac{RIR} applied to speech and far-end differ less than 100~ms in terms of $T_{60}$. We use a 20\% chance of up to 2 echo path changes within a 20~s sequence. The \acp{EIR} are augmented with varying delays up to 0.5~s, the direct path energy is modified with a positively biased gain distribution $\mathcal{N}(12, 5)$~dB, and random bandpass filtering. Finally, the echo signal is added with a \ac{SER} of $\mathcal{N}(0, 10)$~dB.
	Lastly, the signal levels are scaled such that the microphone signals are distributed with $\mathcal{N}(-26, 10)$~dBFS.
	
	The training is monitored using a synthetic validation set of 300 files with 30~s length created in a similar way as the training data. For speech data, we used DAPS, far-end signals are from CommonVoice, and noise from QUT. 
	Training is controlled on the heuristic validation metric
	\begin{equation}
		\label{eq:val_metric}
		\mathcal{V} = nSIG + 0.1\, nOVL + 0.5\, AEC_O + 0.1\, AEC_E
	\end{equation}
	The learning rate is dropped by a factor of 0.5 when $\mathcal{V}$ does not improve for 20 epochs. The final model is chosen on the best $\mathcal{V}$.

	\section{Results}
	
	\subsection{Experimental setup}
	The audio processing and features are implemented in 16~kHz sampling rate using \ac{STFT} with 50\% overlapping 20~ms square-root Hann windows. The feature and loss compression factor is $c=0.3$. As suggested in \cite{Braun2022}, the spectral losses are implemented using a \ac{STFT} with 75\% overlapping 64~ms Hann windows. The deep filter \eqref{eq:mic-enhancement} uses $K\!=\!1$ neighbor frequencies and $L\!=\!2$ past frames.
	The loss terms are weighted with $\alpha\!=\!0.3$, $\beta\!=\!1$, $\gamma_\text{min}\!=\!0.05$, and $\eta\!=\!10^{-5}$.
	One training epoch is defined as 50,400 sequences. The networks are trained with a batch size of 120 for about 500 epochs, resulting in pseudo-unique training data of $\sim$16 years due to random data augmentation.
	All CRUSE models, including the blocks used for DAEC and NRES in Fig.~\ref{fig:twostage-diag}, are of the same size with encoder filters [32,64,64,\textbf{64}] with two exceptions for a larger parameterization of CRUSE-AEC with [32,64,128,\textbf{128}] and a smaller version of the DAEC module with [32,32,32,\textbf{32}]. The model size is indicated by the \textbf{last} filter number.
	
	\subsection{Test data and metrics}
	Results are reported on the 2nd AEC challenge blind test set \cite{Cutler2021} comprising 800 real consumer device recordings with categories near-end, far-end, and double-talk to evaluate \ac{AEC} performance, and on the 3rd DNS challenge \cite{Reddy2021a} blind set comprising 600 real device recordings in challenging noise conditions to evaluate \ac{NS} performance. The \ac{NS} performance is evaluated using the DNSMOS model \cite{Reddy2021b}, which non-intrusively predicts MOS values for signal quality (nSIG), background noise (nBAK) and overall (nOVL) 
	following the ITU-T P.835 standard. The AECMOS model \cite{Purin2022} predicts the echo degradation MOS score $AEC_E$ and other degradation MOS $AEC_O$.
	
	\subsection{Baselines}
	We introduce several baselines as important points of reference. For external reference, we show AECMOS results on the 2nd AEC challenge dataset from the top performing submission (ERCESI)\footnote{https://www.microsoft.com/en-us/research/academic-program/acoustic-echo-cancellation-challenge-interspeech-2021/results/}.
	Secondly, we use a state-of-the-art linear AEC, specifically the \ac{STFT} domain state-space algorithm for AEC described in \cite{LuisValero2019}. The linear AEC is implemented with 75\% overlapping 40~ms windows and an adaptive filter length of 200~ms. Thirdly, we use a model trained on \ac{NS} only (CRUSE-NS-64) to provide a point of reference on a noise suppression dataset, and post-process AEC only models also with the \ac{DNS} model for fair comparison. The CRUSE-NS model \cite{Braun2022} is trained similarly as the \ac{AEC} models on the loss given in \eqref{eq:overall_loss} without the echo term, and does not see echo during training.

	\subsection{Evaluation of overall architecture}
\begin{figure}
		\centering
		\includegraphics[width=\columnwidth,clip,trim=30 10 40 15]{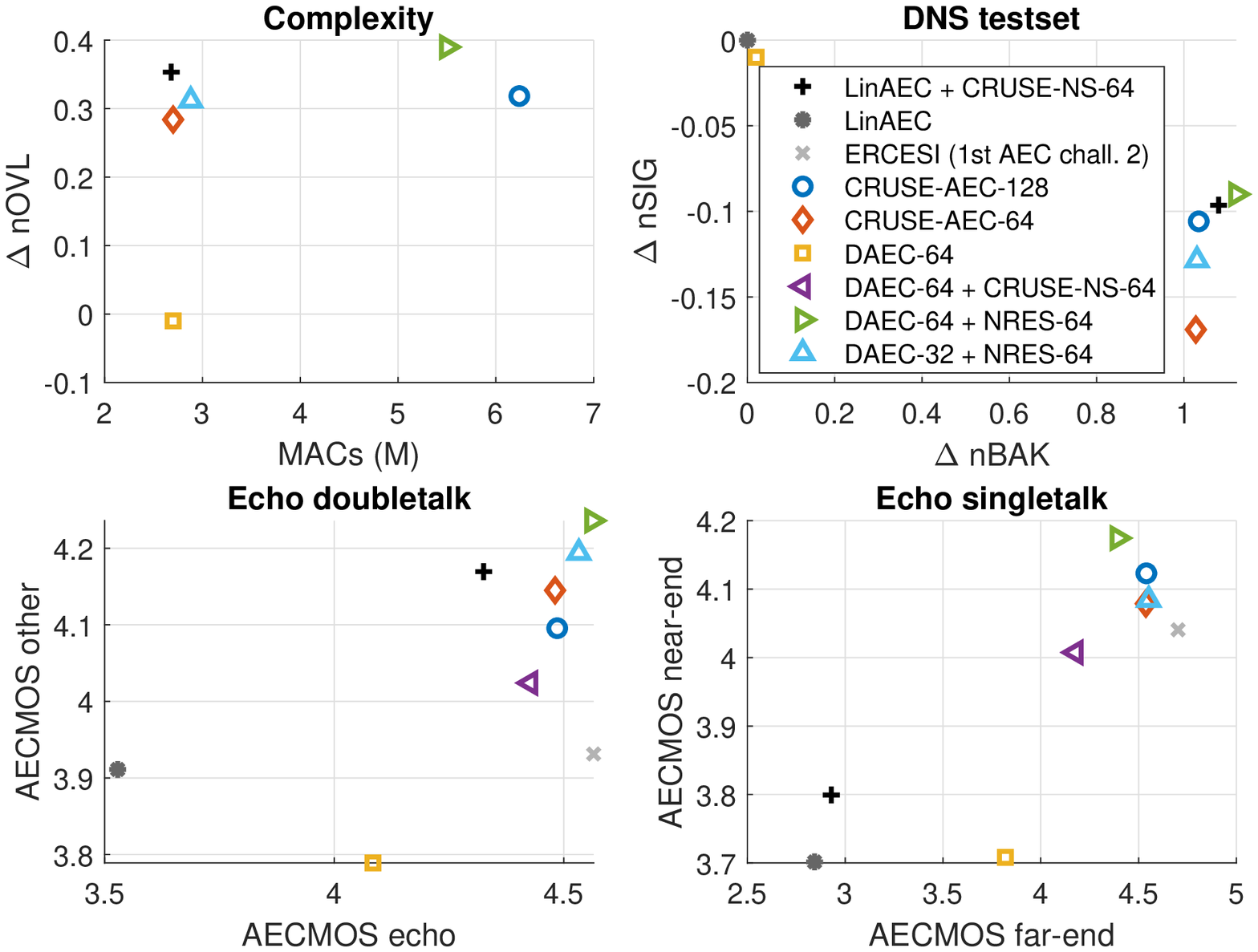}
		\vspace{-.6cm}
		\caption{DNSMOS on 3rd DNS challenge blind set and AECMOS on 2nd AEC challenge blind set. ERCESI not available on NS testset.}
		\label{fig:results-overall}
	\end{figure}

	The overall results are shown in Fig.~\ref{fig:results-overall} in terms of a) inference complexity in \ac{MAC} operations vs.\ overall noise suppression improvement, b) improvement of signal distortion vs.\ noise suppression, c) echo double-talk performance in terms of echo removal vs.\ other (speech) degradation, and d) the single-talk echo tradeoff for far-end suppression vs.\ near-end speech quality. 
	
	%
	Note that for \emph{LinAEC + CRUSE-NS-64}, the complexity is for the NS model only and \emph{LinAEC} is omitted from the top plots as it has no effect on the NS testset. We can see that the NS-only model achieves good overall noise suppression at 2.7M MACs and good nBAK with only little nSIG degradation. Interestingly, the linear AEC is greatly improved for doubletalk by chaining the NS model.
	
	The single-stage AEC model CRUSE-AEC-64 which is same size as the \ac{DNS} only model CRUSE-NS-64 shows decent echo performance, but performs significantly worse on the DNS testset than CRUSE-NS-64, especially for \emph{nSIG}, as it has to learn two tasks. The scaled up CRUSE-AEC-128 with 2.5x complexity increase achieves increased echo performance, but still shows some NS performance degradation compared to CRUSE-NS-64. The external reference model from ERCESI shows slightly more single-talk echo suppression with however a large degradation on the near-end/other metric, indicating either more speech distortions or less noise suppression.
	
	
	Evaluating only the first module of the proposed two-stage model, i.\,e.\, the \ac{DAEC} output $e(t)$, we observe some desired effects: The DAEC-64 model removes no noise, therefore also causing no $nSIG$ degradation, which can be beneficial. It removes the echo already almost completely as indicated by an echo MOS score $>4$, while lagging in the echo other/near-end categories due to passing through noise. A significant improvement is obtained by chaining the \ac{DAEC} model with CRUSE-NS, which are separately trained (CRUSE-NS never saw echo). The performance of this sub-optimal chain is obviously lower than the joint task models CRUSE-AEC and the full two-stage DAEC+NRES models.
	
	The full proposed two-stage system \emph{DAEC-64\,+\,NRES-64} achieves better NS and echo performance except for the single-talk far-end metric than the single-stage large \emph{CRUSE-AEC-128} model at similar complexity. 
	We test the hypothesis that the semi-supervised task of echo cancellation may be easier to learn for neural networks than the blind task of noise suppression by spending less complexity for the \ac{DAEC} module to obtain more efficient models. For \emph{DAEC-32\,+\,NRES-64} the \ac{DAEC} module is down-sized while keeping the NRES module the same size as the DNS model. While the down-sized \emph{DAEC-32\,+\,NRES-64} shows minor degradations in echo and NS performance, it significantly outperforms \emph{CRUSE-AEC-64 } in nSIG and double-talk at similar complexity.


	
	\subsection{Loss ablation for two-stage model}
	An ablation study shown in Tab.~\ref{tab:ablation} provides insights into the proposed loss.
	%
	%
	The first row is the proposed training method with asymmetric loss term \eqref{eq:asymloss} weighted with $\alpha\!=\!0.5$ and the adaptive echo loss weighting \eqref{eq:res_echo_weight}. Dropping the asymmetric loss term by setting $\alpha\!=\!0$ leads to increased \ac{NS} and singletalk echo reduction at the expense of signal distortion in terms of nSIG and AECMOS other/near-end. 
	When replacing the adaptive echo loss weighting \eqref{eq:res_echo_weight} with a fixed weight, we found $\gamma\!=\!\gamma_\text{min}\!=\!0.05$ to be optimal, we can see a drop in all metrics. The DAEC and NRES modules were trained all from scratch here. The performance drop can be remedied by pre-training the DAEC module on the echo loss \eqref{eq:echoloss}, and then finetuning the whole DAEC+NRES pipeline on the overall loss \eqref{eq:overall_loss} with the fixed echo loss weight $\gamma\!=\!0.05$. During finetuning we kept updating the weights of both DAEC and NRES as freezing the DAEC weights led to significantly worse results. While with pretrained DAEC, the fixed echo loss achieves similar results to the proposed adaptive echo loss \eqref{eq:res_echo_weight}, the laborsome pretraining step can be avoided.
	\begin{table}[tb]
		\scriptsize
		\centering
		\begin{tabular}{p{.7cm}p{.45cm}p{.35cm}||p{.3cm}p{.35cm}|p{.3cm}p{.3cm}|p{.38cm}p{.38cm}p{.38cm}} 
		\toprule
		& & & \multicolumn{2}{c|}{doubletalk} & \multicolumn{2}{c|}{singletalk} & \multicolumn{3}{c}{DNS} \\
		pretrain & $\gamma$ & $\alpha$ & echo & other & FE & NE & nSIG & nBAK & nOVL \\\midrule
		-- & eq.~\eqref{eq:res_echo_weight} & 0.5 &  \bf{4.55}	& \bf{4.25}	& 4.35	& \bf{4.18} & \bf{3.68}	& 4.26	& \bf{3.50} \\	
		-- & eq.~\eqref{eq:res_echo_weight} & 0 &  \bf{4.55} & 4.22 & \bf{4.39} & 4.17 & 3.65 & \bf{4.27} & 3.49 \\	
		-- & 0.05 & 0.5 &  \bf{4.55} &  4.24 & 4.29 & 4.17 & 3.65 & 4.23 & 3.47 \\	
		DAEC & 0.05 & 0.5 &  \bf{4.55} & \bf{4.2}5 & 4.35 & 4.17 & \bf{3.68} & 4.26 & \bf{3.50} \\
		\bottomrule
	\end{tabular}
	\caption{Ablation for loss and training strategy for DAEC+NRES.}
	\label{tab:ablation}
\end{table}
%


\subsection{Challenging scenario analysis}
\label{sec:breakdown-analysis}
We investigate further how AEC algorithms deal with specific practical challenges that violate basic assumptions of linear AECs. Specifically, we show results of AECMOS on the far-end singletalk subset in the following categories, which were annotated by a human expert listener: \emph{clean, noise, volume changes, long delays, AEC (not deactivated on-device pre-processing), non-linear distortions (loudspeaker clipping), glitches, other}.

\begin{figure}[tb]
	\centering
	\includegraphics[width=\columnwidth,clip,trim=60 0 60 0]{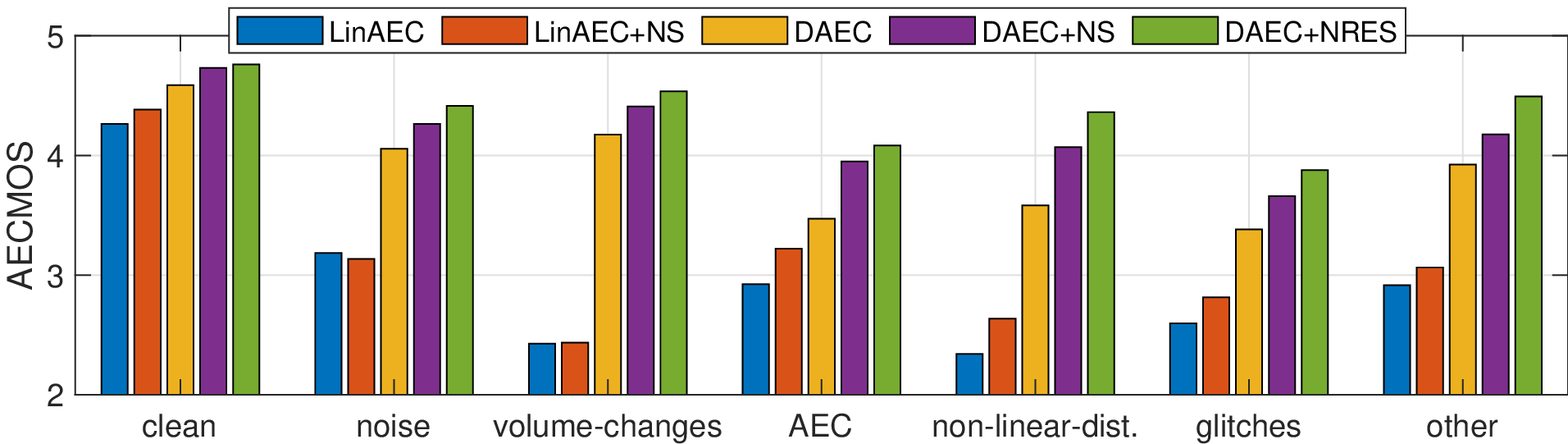}
	\vspace{-.6cm}
	\caption{Breakdown of far-end singletalk for challenging scenarios.}
	\label{fig:farend-analysis}
\end{figure}
We can see that the linear AEC struggles in all cases except \emph{clean} by showing significantly lower AECMOS than the \acp{DNN}. While the gap of the DAEC to the DAEC+NRES is small for clean, noise, volume-changes, it is interesting that
there is still a significant performance gap for the categories AEC, non-linear-distortions, glitches, other.
This suggests that the DAEC module is faster reacting and adapting than LinAEC, but still is not sufficient to remove highly non-linear, non-stationary and other uncommon artifacts, where a model trained on a speech enhancement task (NS) or the jointly trained NRES module can provide large benefits.

\section{Conclusions}
We showed that separating the tasks for \ac{AEC} and \ac{NRES} into individual modules can achieve more efficient models at lower speech distortion and better double-talk performance than using the same \ac{DNN} model at adjusted capacity for both tasks. We propose an adaptive loss that guides the model training to focus first on the \ac{AEC} module and optimizes the overall system later in the training end-to-end. We show an analysis of difficult scenarios, where the \ac{DAEC} significantly outperforms a linear \ac{AEC} with faster reaction times and higher robustness to noise and non-linearities. There are however some categories, where the \ac{DAEC} removes echo components only partially, and the following \ac{NRES} module shows significant additional improvements.
Practical advantages of the decoupled two-stage system are that the \ac{AEC} can be switched off when not needed, \ac{AEC} and \ac{NRES} network architectures can be tuned individually in size and architecture, and the echo signal is obtained inherently for analysis or re-use in other applications. Future work is required to optimize the network module architectures to the individual tasks.

\balance
\bibliographystyle{IEEEbib}
\bibliography{./sapref.bib}

\begin{thebibliography}{10}

\bibitem{Hansler2004}
E.~H{\"{a}}nsler and G.~Schmidt,
\newblock {\em Acoustic Echo and Noise Control: A pracical Approach},
\newblock Wiley, New Jersey, USA, 2004.

\bibitem{Tan2020}
K.~{Tan} and D.~{Wang},
\newblock ``Learning complex spectral mapping with gated convolutional
  recurrent networks for monaural speech enhancement,''
\newblock vol. 28, pp. 380--390, 2020.

\bibitem{Fedorov2020}
Igor Fedorov, Marko Stamenovic, Carl Jensen, Li-Chia Yang, Ari Mandell, Yiming
  Gan, Matthew Mattina, and Paul~N. Whatmough,
\newblock ``{TinyLSTMs}: Efficient neural speech enhancement for hearing
  aids,''
\newblock in {\em Proc. Interspeech Conf.}, 2020.

\bibitem{Braun2021a}
S.~Braun, H.~Gamper, C.~K.~A. Reddy, and I.~Tashev,
\newblock ``Towards efficient models for real-time deep noise suppression,''
\newblock in {\em Proc. {IEEE} Intl. Conf. on Acoustics, Speech and Signal
  Processing (ICASSP)}, 2021.

\bibitem{Cutler2021}
Ross Cutler, Ando Saabas, Tanel Parnamaa, Markus Loide, Sten Sootla, Marju
  Purin, Hannes Gamper, Sebastian Braun, Karsten Sorensen, Robert Aichner, and
  Sriram Srinivasan,
\newblock ``Interspeech 2021 acoustic echo cancellation challenge,''
\newblock in {\em Proc. Interspeech Conf.}, 2021.

\bibitem{Carbajal2020}
G.~Carbajal, R.~Serizel, E.~Vincent, and E.~Humbert,
\newblock ``Joint {NN}-supported multichannel reduction of acoustic echo,
  reverberation and noise,''
\newblock {\em IEEE/ACM Transactions on Audio, Speech, and Language
  Processing}, vol. 28, pp. 2158--2173, 2020.

\bibitem{Halimeh2019}
M.~M. Halimeh, C.~Huemmer, and W.~Kellermann,
\newblock ``A neural network-based nonlinear acoustic echo canceller,''
\newblock {\em IEEE Signal Processing Letters}, vol. 26, no. 12, pp.
  1827--1831, 2019.

\bibitem{Ivry2021}
Amir Ivry, Israel Cohen, and Baruch Berdugo,
\newblock ``Nonlinear acoustic echo cancellation with deep learning,'' arXiv
  preprint https://arxiv.org/abs/2106.13754, 2021.

\bibitem{Zhang2019}
H.~Zhang, K.~Tan, and D.~Wang,
\newblock ``Deep learning for joint acoustic echo and noise cancellation with
  nonlinear distortions,''
\newblock in {\em Proc. Interspeech Conf.}

\bibitem{Peng2021}
R.~Peng, L.~Cheng, C.~Zheng, and X.~Li,
\newblock ``Acoustic echo cancellation using deep complex neural networkwith
  nonlinear magnitude compression and phase information,''
\newblock in {\em Proc. Interspeech Conf.}

\bibitem{Westhausen2021}
Nils~L. Westhausen and Bernd~T. Meyer,
\newblock ``Acoustic echo cancellation with the dual-signal transformation
  {LSTM} network,''
\newblock in {\em Proc. {IEEE} Intl. Conf. on Acoustics, Speech and Signal
  Processing (ICASSP)}, 2021, pp. 7138--7142.

\bibitem{Valin2021}
J.~M. Valin, S.~Tenneti, K.~Helwani, U.~Isik, and A.~Krishnaswamy,
\newblock ``Low-complexity, real-time joint neural echo control and speech
  enhancement based on percepnet,''
\newblock in {\em Proc. {IEEE} Intl. Conf. on Acoustics, Speech and Signal
  Processing (ICASSP)}, 2021, pp. 7133--7137.

\bibitem{Franzen2021}
J.~Franzen, E.~Seidel, and T.~Fingscheidt,
\newblock ``{AEC} in a netshell: on target and topology choices for {FCRN}
  acoustic echo cancellation,''
\newblock in {\em Proc. {IEEE} Intl. Conf. on Acoustics, Speech and Signal
  Processing (ICASSP)}, 2021, pp. 156--160.

\bibitem{Seidel2021}
E.~Seidel, J.~Franzen, M.~Strake, and T.~Fingscheidt,
\newblock ``Y$^2$-net {FCRN} for acoustic echo and noise suppression,''
\newblock in {\em Proc. Interspeech Conf.}, Brno, Czechia, 2021.

\bibitem{Li2021}
A.~Li, W.~Liu, X.~Luo, G.~Yu, C.~Zheng, and X.~Li,
\newblock ``A simultaneous denoising and dereverberation framework with target
  decoupling,''
\newblock in {\em Proc. Interspeech Conf.}, 2021.

\bibitem{Ivry2021a}
A.~Ivry, I.~Cohen, and B.~Berdugo,
\newblock ``Deep residual echo suppression with a tunable tradeoff between
  signal distortion and echo suppression,''
\newblock in {\em Proc. {IEEE} Intl. Conf. on Acoustics, Speech and Signal
  Processing (ICASSP)}, 2021, pp. 126--130.

\bibitem{Mack2019}
W.~{Mack} and E.~A.~P. {Habets},
\newblock ``Deep filtering: Signal extraction and reconstruction using complex
  time-frequency filters,''
\newblock {\em {IEEE} Signal Process. Lett.}, pp. 1--5, 2019.

\bibitem{Braun2022}
S.~Braun and H.~Gamper,
\newblock ``Effect of noise suppression losses on speech distortion and {ASR}
  performance,''
\newblock in {\em Proc. {IEEE} Intl. Conf. on Acoustics, Speech and Signal
  Processing (ICASSP)}, Singapore, 2022.

\bibitem{Li2021a}
A.~Li, C.~Zheng, R.~Penga, and X.~Li,
\newblock ``On the importance of power compression and phase estimation in
  monaural speech dereverberation,''
\newblock {\em JASA express letters}, vol. 1, no. 014802, 2021.

\bibitem{Braun2021b}
S.~Braun and I.~Tashev,
\newblock ``A consolidated view of loss functions for supervised deep
  learning-based speech enhancement,''
\newblock in {\em Intl. Conf. on Telecomm. and Sig. Proc. (TSP)}, 2021.

\bibitem{Wang2020a}
Q.~Wang, I.~Lopez~Moreno, M.~Saglam, K.~Wilson, A.~Chiao, R.~Liu, Y.~He, W.~Li,
  J.~Pelecanos, M.~Nika, and A.~Gruenstein,
\newblock ``Voicefilter-lite: Streaming targeted voice separation for on-device
  speech recognition,''
\newblock in {\em Proc. Interspeech Conf.}, 2020.

\bibitem{Reddy2021a}
C.~K.~A. Reddy, H.~Dubey, K.~Koishida, A.~Nair, V.~Gopal, R.~Cutler, S.~Braun,
  R.~Gamper, H.~Aichner, and S.~Srinivasan,
\newblock ``{INTERSPEECH} 2021 deep noise suppression challenge:,''
\newblock in {\em Proc. Interspeech Conf.}, 2021.

\bibitem{Ephrat2018}
A.~Ephrat, I.~Mosseri, O.~Lang, T.~Dekel, K.~Wilson, A.~Hassidim, W.~T.
  Freeman, and M.~Rubinstein,
\newblock ``Looking to listen at the cocktail party: A speaker-independent
  audio-visual model for speech separation,''
\newblock {\em ACM Trans. Graph.}, vol. 37, no. 4, July 2018.

\bibitem{Chung2018}
J.~S. Chung, A.~Nagrani, and A.~Zisserman,
\newblock ``Voxceleb2: Deep speaker recognition,''
\newblock in {\em Proc. Interspeech Conf.}, 2018.

\bibitem{Snyder2015}
David Snyder, Guoguo Chen, and Daniel Povey,
\newblock ``{MUSAN}: {A} {M}usic, {S}peech, and {N}oise {C}orpus,'' 2015,
\newblock arXiv:1510.08484v1.

\bibitem{Reddy2021b}
C.~Reddy, V.~Gopal, and R.~Cutler,
\newblock ``{DNSMOS}: A non-intrusive perceptual objective speech quality
  metric to evaluate noise suppressors,''
\newblock in {\em Proc. {IEEE} Intl. Conf. on Acoustics, Speech and Signal
  Processing (ICASSP)}, October 2021.

\bibitem{Purin2022}
M.~Purin, S.~Sootla, M.~Sponza, A.~Saabas, and R.~Cutler,
\newblock ``{AECMOS}: A speech quality assessment metric for echo impairment,''
\newblock in {\em Proc. {IEEE} Intl. Conf. on Acoustics, Speech and Signal
  Processing (ICASSP)}, 2022, pp. 901--905.

\bibitem{LuisValero2019}
M.~Luis~Valero and E.~A.~P. Habets,
\newblock ``Low-complexity multi-microphone acoustic echo control in the
  short-time fourier transform domain,''
\newblock {\em IEEE/ACM Transactions on Audio, Speech, and Language
  Processing}, vol. 27, no. 3, pp. 595--609, 2019.

\end{thebibliography}

\end{document}